\documentclass{iau}
\usepackage{graphicx}
\usepackage{hyperref}

\newcommand{\msim}{\raisebox{-.4ex}{$\stackrel{>}{\scriptstyle \sim}$}}
\def \etal   {\hbox{et~al.\/}}

\title[Clumped stellar winds in HMXBs] 
{Clumped stellar winds in supergiant high-mass X-ray binaries}

\author[Oskinova, Feldmeier, Kretschmar]   
{L. M. Oskinova$^1$, A. Feldmeier$^1$, P. Kretschmar$^2$}
\affiliation{$^1$Institute for Physics and Astronomy, University of Potsdam, 14476 Potsdam, Germany,
\\ email: {\tt lida@astro.physik.uni-potsdam.de} \\ 
$^2$European Space Astronomy Centre (ESA/ESAC), Science Operations Department, 
Villanueva de la Ca\~{n}ada, Madrid, Spain}

\pubyear{2012}
\volume{290}  
\jname{Feeding compact objects: Accretion on all scales}
\editors{C.M. Zhang, T. Belloni, M. M\'endez \& S.N. Zhang, eds.}

\begin{document}

\maketitle

\begin{abstract}
The clumping of massive star winds is an established paradigm, which is 
confirmed by multiple lines of evidence and is supported by stellar wind theory.
We use the results from time-dependent hydrodynamical models of the instability 
in the line-driven wind of a massive supergiant star to derive the time-dependent
accretion rate on to a compact object in the Bondi-Hoyle-Lyttleton approximation. 
The strong density and velocity fluctuations in the wind result in strong 
variability of the synthetic X-ray light curves. Photoionization of inhomogeneous 
winds is different from the photoinization of smooth winds. The degree of ionization 
is affected by the wind clumping. The wind clumping must also be 
taken into account when comparing the observed and model spectra of the photoionized 
stellar wind. 

\end{abstract}

\firstsection 
\section{Clumping of massive star winds}
Massive luminous OB-type stars possess strong stellar winds. The winds
are fast, with typical velocities up to 2500\,km\,s$^{-1}$, and dense,
with mass-loss rates $\dot{M}\msim 10^{-7}\,M_\odot$\,yr$^{-1}$. The driving
mechanism for the mass-loss from OB stars has been identified with
radiation pressure on spectral lines (Castor \etal\ 1975).  Early on
Lucy \& Solomon (1970) suggested that the stationary solution for a
line-driven wind is unstable. 
Lucy \& White (1980) proposed that the winds break up into a population of
dense blobs. Feldmeier \etal\ (1997a,b) used hydrodynamic simulations to
model the evolution of wind instabilities and found that the winds are 
non-stationary and inhomogeneous. 
The inhomogeneity affects stellar wind diagnostics. Stewart \& Fabian (1981)
used \emph{Einstein} spectra of O-stars to study the transfer of
X-rays through a uniform stellar wind.  They found that the mass-loss 
rate derived from X-ray spectra is lower by factor of a few than the 
mass-loss rates obtained from fitting the H$\alpha$ line.  As most 
plausible explanation for this discrepancy they suggested the neglect 
of the clumping in O-star winds.  A theory of X-ray transfer in clumped 
winds that accounts for clumps of any shapes and optical depths was 
developed by Feldmeier \etal\ (2003) and Oskinova \etal\ (2004). Indeed, 
accounting for wind clumping allows to consistently model the UV/optical 
and the X-ray spectra of single O-stars (Oskinova \etal\ 2006, 2007). These 
conclusions are confirmed by full 3-D radiative-transfer models of stellar 
winds (\v{S}urlan \etal\ 2012).

The mass-loss rates of massive stars are in good agreement with observed 
X-ray fluxes from high-mass X-ray binaries (HMXBs). HMXBs are the products 
of binary star evolution. These systems consist  of an early-type massive 
star and a compact companion, neutron star (NS) or black hole (BH), on a 
close orbit (Iben \etal\ 1995). Accretion of stellar  wind onto the compact 
companion powers the high X-ray luminosity of 
$\sim 10^{33}$\,--\,$10^{39}$\,erg\,s$^{-1}$. HMXBs can be divided
in subclasses depending on the spectral type of the donor star --
supergiant (SG) or Be-star.

\section{Effect of wind-clumping for HMXBs}

\begin{figure}[t]
\begin{center}
 \includegraphics[width=3.4in]{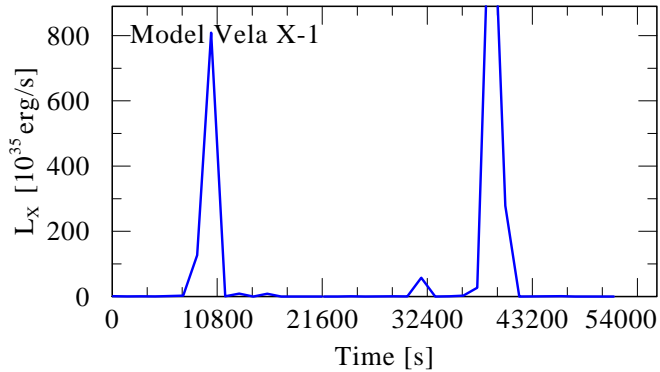} 
  \caption{Synthetic X-ray light curves 
for the Bondi–Hoyle accretion of a non-stationary 
stellar wind on to a NS with $m_{\rm X}= 1.4\,M_\odot$. 
Parameters of Vela X-1 were used in the simulation.} 
    \label{fig1}
 \end{center}
\end{figure}

We employ the results of hydrodynamical simulations of the line driven
stellar wind presented in Feldmeier \etal\ (1997a,b). These
hydrodynamic simulations derive the dynamic and thermal structure of
stellar winds from the basic underlying physical principle, namely the
acceleration of stellar wind by the line scattering of stellar UV
photons.

The time dependent hydrodynamic simulations  provide 
absolute values for the wind density and velocity as functions of radius 
and time. We use the predicted density and velocity field to compute synthetic X-ray 
light curves. An example of a light-curve is shown in Fig.\,1. The variability 
is very large because of the strong variations in the wind velocity (shocks). 
The highly non-uniform wind density also contributes to the variability of accretion rate.
The variations in the model light-curve are much larger than even strongest variability
observed in HMXBs. We are prompted to conclude that simple Bondi-Hoyle
accretion with clumped winds is not a sufficient explanation
of variations in wind-accreting X-ray binaries.

Oskinova \etal\ (2012) showed that the photoionization of the  wind by the 
radiation from accreting object is strongly affected  by the wind clumping. 
In the optically thin case, the photoionization parameter has to be reduced 
by a factor $\chi$  compared to the smooth wind model, where $\chi$ is the 
wind inhomogeneity parameter.

Thus, wind clumping affects the accretion regime, the photoionization, and 
the wind opacity for the X-rays.

\end{document}